\documentclass[letterpaper,twocolumn,10pt]{article}
\usepackage{usenix-2020-09}

\usepackage{microtype}
\usepackage{graphicx}
\usepackage{booktabs} %
\usepackage{multirow}

\usepackage{algorithm}
\usepackage{algpseudocode}
\usepackage{authblk}

\usepackage{enumitem}
\usepackage{xspace}
\usepackage{amsmath}
\usepackage{amssymb}
\usepackage{amsfonts}
\usepackage[svgnames,table,dvipsnames]{xcolor}
\usepackage{url}

\usepackage{hyperref}
\hypersetup{
    colorlinks,
    linkcolor={blue!50!black},
    citecolor={red!50!black},
    urlcolor={blue!80!black}
}

\usepackage{ifthen}

\usepackage{subcaption}

\usepackage{cleveref}
\crefformat{section}{\S#2#1#3}
\crefmultiformat{section}{\S#2#1#3}{ and~\S#2#1#3}{, \S#2#1#3}{ and~\S#2#1#3} %

\newcommand{\sysname}{\textsc{Revati}\xspace}

\newcommand\blfootnote[1]{%
  \begingroup
  \renewcommand\thefootnote{}\footnote{#1}%
  \addtocounter{footnote}{-1}%
  \endgroup
}

\newcommand{\vidur}{Vidur\xspace}

\definecolor{codegreen}{rgb}{0,0.6,0}

\newcommand{\greencheck}{\textcolor{codegreen}{\checkmark}}
\newcommand{\redcross}{\textcolor{red}{$\times$}}

\newcommand{\timekeeper}{\textit{Timekeeper}\xspace}
\newcommand{\actors}{\textit{Actors}\xspace}
\newcommand{\observers}{\textit{Observers}\xspace}
\newcommand{\actor}{\textit{Actor}\xspace}

\newcommand{\mycaption}[2]{\caption{\textbf{#1} {#2}}}

\newcommand{\vheading}[1]{\vspace{0.05in}\noindent\textbf{#1.}}

\newcommand{\begincompactitemize}{\begin{itemize}[noitemsep,topsep=0pt,parsep=0pt,partopsep=0pt]}

\newcommand{\todo}[1]{}

\newcommand{\llamaS}{Llama-3.1-8B\xspace}
\newcommand{\llamaL}{Llama-3.1-70B\xspace}
\newcommand{\qwenMM}{Qwen3-30B-A3B\xspace}

\newboolean{publicversion}
\setboolean{publicversion}{true}  %

\ifthenelse{\boolean{publicversion}}{
    \newcommand{\grumbler}[3]{}
    \newcommand{\amey}[1]{}
    \newcommand{\alexey}[1]{}
    \newcommand{\anirudha}[1]{}
    \newcommand{\sirish}[1]{}
    \newcommand{\mayank}[1]{}
    \newcommand{\sukrit}[1]{}
    \newcommand{\srinivas}[1]{}
    \newcommand{\elton}[1]{}
    \newcommand{\souradeep}[1]{}
    \newcommand{\authorcomment}[1]{}
}
{
     \newcommand{\grumbler}[3]{}
    \newcommand{\amey}[1]{}
    \newcommand{\alexey}[1]{}
    \newcommand{\anirudha}[1]{}
    \newcommand{\sirish}[1]{}
    \newcommand{\mayank}[1]{}
    \newcommand{\sukrit}[1]{}
    \newcommand{\srinivas}[1]{}
    \newcommand{\elton}[1]{}
    \newcommand{\souradeep}[1]{}
    \newcommand{\authorcomment}[1]{}
}

\usepackage[margin=1in]{geometry}
 
\setlist[itemize]{itemsep=1pt, topsep=1pt}

\begin{document}

\title{\sysname: Transparent GPU-Free Time-Warp Emulation for LLM Serving}

\author{Amey Agrawal$^*$}
\author{Mayank Yadav$^*$}
\author{Sukrit Kumar$^\dagger$}
\author{Anirudha Agrawal$^\dagger$}
\author{Garv Ghai$^\dagger$}
\author{\hspace{10em}Souradeep Bera}
\author{Elton Pinto}
\author{Sirish Gambhira}
\author{Mohammad Adain}
\author{Kasra Sohrab}
\author{\hspace{10em}Chus Antonanzas}
\author{Alexey Tumanov}

\affil{Georgia Institute of Technology}

\maketitle

\blfootnote{$^*$$^\dagger$ Denote equal contribution levels. Revati is a mythological character who features in one of the earliest fictional tales of time dilation.}

\begin{abstract}
    Deploying LLMs efficiently requires testing hundreds of serving configurations, but evaluating each one on a GPU cluster takes hours and costs thousands of dollars. Discrete-event simulators are faster and cheaper, but they require re-implementing the serving system's control logic --- a burden that compounds as frameworks evolve.

    We present \sysname{}, a \textit{time-warp emulator} that enables performance modeling by directly executing real serving system code at simulation-like speed. The system intercepts CUDA API calls to virtualize device management, allowing serving frameworks to run without physical GPUs. Instead of executing GPU kernels, it performs \textit{time jumps} -- fast-forwarding virtual time by predicted kernel durations. We propose a coordination protocol that synchronizes these jumps across distributed processes while preserving causality. On vLLM and SGLang, \sysname{} achieves $<$5\% prediction error across multiple models and parallelism configurations, while running 5-17$\times$ faster than real GPU execution.    
\end{abstract}

\section{Introduction}

LLM inference increasingly dominates AI costs. To manage these costs, operators need to tune a vast set of configuration knobs, including, various parallelization strategies, caching policies, and scheduling algorithms \cite{vidur,maya}. These knobs must be tuned for each model, workload and hardware platform as they evolve over time. Finding the optimal deployment configuration through trial-and-error end-to-end evaluation is prohibitively slow and expensive. Evaluating even a single configuration on a GPU cluster can take several hours and cost thousands of dollars in compute \cite{vidur}. This bottlenecks both production deployment and systems research.

\begin{figure}[t]
\centering
\includegraphics[width=0.75\columnwidth]{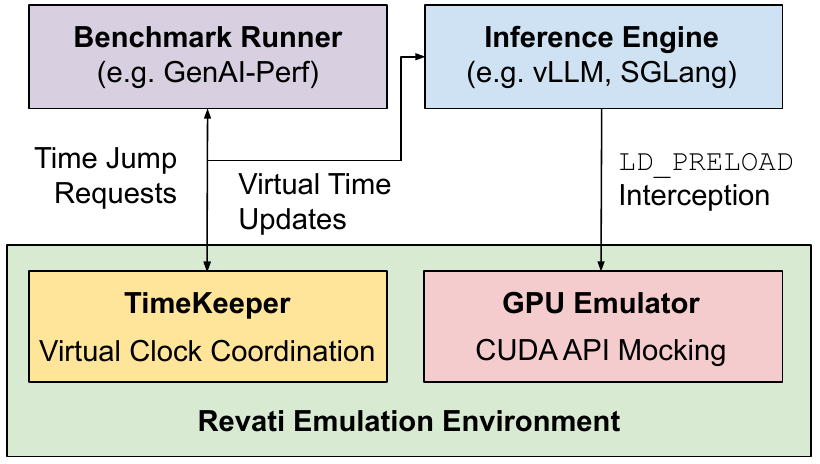}
\caption{ Discrete-event simulators must re-implement the entire control-flow and scheduling logic of serving systems, as a result, the rapid advancements in LLM inference render these simulators perpetually outdated. \sysname{} \textbf{\textit{entirely eliminates}} this problem by \textbf{\textit{directly}} running the serving systems in a emulated  environment.}
\end{figure}

\vheading{Limitations of Current Approaches} To address this challenge, the community has developed discrete-event simulators ~\cite{vidur, llmservingsim, apex} that can cheaply model serving performance. Simulators offer compelling benefits: they run orders of magnitude faster than real execution and are easy to modify for prototyping new policies and optimization. However, they require re-implementing all the core system components including the scheduler, memory allocator, and request dispatcher, etc. This approached worked when systems were simple: modeling vLLM's original continuous batching scheduler took less than 150 lines to implement in \vidur~\cite{vidur}. However, today LLM engines include complex features like prefill-decode disaggregation \cite{splitwise, distserve, mooncake, nvidia-dynamo,cheng2025lmcache} and distributed prefix caching \cite{nvidia-dynamo, cheng2025lmcache} with implementations spanning tens of thousands of lines of code. Moreover, with thousands of commits per year from hundreds of contributors, inference serving systems are evolving at a rapid pace, causing simulators to perpetually lag behind with surmounting maintenance burden.

\vheading{Key Insight} Serving systems spend most of their execution time waiting for GPU computation, \textit{not} making control decisions. The CPU control plane (scheduling, batching, memory management) executes in microseconds, while GPU operations take tens of milliseconds accounting for more than 90\% of execution time. Moreover, the control flow logic is largely decoupled from the output of GPU computation. This separation enables a radical alternative: run the \textit{real} serving system while skipping GPU waits through virtual time advancement. We call this \textbf{\textit{Time-Warp Emulation}}.

\vheading{Our Approach} \sysname executes actual serving frameworks like vLLM or SGLang with GPU operations replaced by virtual time jumps. When a worker prepares to execute a batch, it asks: ``How long would this take on H100?''. Using the predicted duration, it requests a \textbf{\textit{time jump}} from a central \timekeeper that coordinates virtual time across all processes. The \timekeeper ensures all processes advance together while preserving causality: if worker A needs 15ms and worker B needs 25ms, the system advances by 15ms first. A CUDA emulation layer transparently intercepts GPU calls, presenting virtual devices to the serving framework while bypassing actual execution. This allows us to directly execute the serving system's code \textit{without need for actual GPUs} in accelerated virtual time.

By running an unchanged control plane, \sysname automatically captures complex system behaviors without requiring manual modeling. Complex features like adaptive batching, multi-tiered prefix cache management, PD disaggregation, etc. work out of the box. \sysname poses a minimal maintenance overhead with less than 25 lines of change required to onboard a serving system. This combination of speed, cost-effectiveness, and ease of adoption enables rapid exploration of optimal deployment configurations and accelerates both the development and adoption of state-of-the-art LLM inference techniques.

\begin{figure}[t]
\centering
\includegraphics[width=\columnwidth]{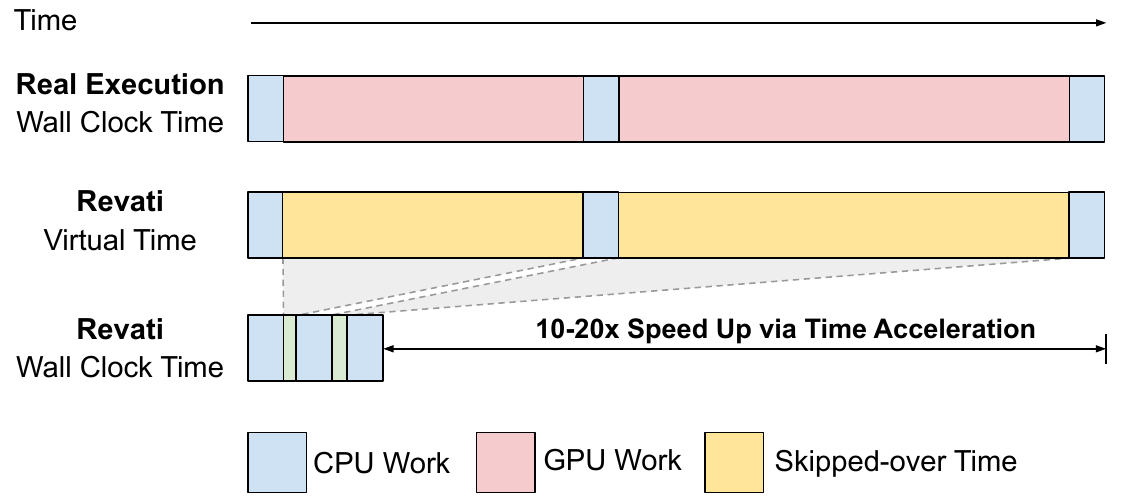}
\mycaption{Time Accelerated Emulation.}{LLM inference engines try to maximize the GPU computation (red); while keeping the CPU overhead (blue) as minimal as possible. \sysname{} exploits this property by skipping GPU operations in wall clock time while preserving the causal semantics of the application using virtual time semantics. This allows us to obtain an order of magnitude acceleration without sacrificing on fidelity.}
\end{figure}

\begin{table}[t]
\centering
\small
\setlength{\tabcolsep}{2pt}
\resizebox{\columnwidth}{!}{%
\begin{tabular}{lccccccc}
\toprule
& \textbf{\sysname} & \multicolumn{6}{c}{\textbf{Discrete-Event Simulators}} \\
\cmidrule(lr){3-8}
&  & VD & LS1 & AP & FT & TS & LS2 \\
\midrule
\rowcolor{gray!15} \multicolumn{8}{l}{\textit{System Properties}} \\
Direct System Emulation & \greencheck & \redcross & \redcross & \redcross & \redcross & \redcross & \redcross \\
Serving System Agnostic & \greencheck & \redcross & \redcross & \redcross & \redcross & \redcross & \redcross \\
Minimal Maintenance Overhead & \greencheck & \redcross & \redcross & \redcross & \redcross & \redcross & \redcross \\
\midrule
\rowcolor{gray!15} \multicolumn{8}{l}{\textit{Modeling Domain}} \\
Continuous Batching & \greencheck & \greencheck & \greencheck & \greencheck & \greencheck & \greencheck & \greencheck \\
Chunked Prefill & \greencheck & \greencheck & \greencheck & \greencheck & \greencheck & \greencheck & \greencheck \\
Prefix Caching & \greencheck & \redcross & \redcross & \redcross & \redcross & \greencheck & \greencheck \\
Hierarchical Caching & \greencheck & \redcross & \redcross & \redcross & \redcross & \redcross & \greencheck \\
PD Disaggregation & \greencheck & \redcross & \redcross & \redcross & \greencheck & \greencheck & \greencheck \\
DP Attention & \greencheck & \redcross & \redcross & \redcross & \greencheck & \redcross & \greencheck \\
MoE / Expert Parallel & \greencheck & \redcross & \redcross & \greencheck & \greencheck & \redcross & \greencheck \\
\bottomrule
\end{tabular}%
}
\caption{Comparison of LLM inference modeling approaches. Discrete-event simulators require manual re-implementation of features, creating maintenance burden. Older simulators like Vidur (VD) \cite{vidur}, LLMServingSim (LS1) \cite{llmservingsim}, and APEX (AP) \cite{apex} lack critical modern features. Newer simulators like Frontier (FT) \cite{frontier}, TokenSim (TS) \cite{tokensim}, and LLMServingSim2.0 (LS2) \cite{llmservingsim2} fill these gaps but face similar maintenance challenges as their predecessors.}
\label{tab:comparison}

\end{table}

\vspace{2em}

\noindent This paper makes the following contributions:

\begin{itemize}
\item We present \sysname{}, a time-warp emulator that models serving system performance at simulation-like speed without re-implementing system logic.
\item We develop a virtual time protocol that coordinates time jumps across distributed processes while preserving causality.
\item We evaluate \sysname{} on vLLM and SGLang, achieving $<$5\% prediction error while running 5--17$\times$ faster than real GPU execution.
\end{itemize}

\section{Background \& Motivation}
\label{sec:background}

Performance modeling is essential for the efficient deployment of LLM serving systems because real-world experimentation on GPU clusters is prohibitively expensive. However, existing simulator-based approaches struggle to maintain fidelity as serving systems evolve due to lagging feature parity. This section characterizes the configuration optimization challenge and illustrates why discrete-event simulation fails to meet the needs of modern serving infrastructure.

\subsection{Configuration Optimization Problem}
Modern LLM serving systems expose vast configuration spaces. Consider deploying Qwen3-235B-A22B \cite{qwen3}: operators must select the tensor parallelism (TP) degree (1-8), number of pipeline parallelism (PP) stages (1-8), expert parallelism (EP) degree (1-8), maximum batch size (1-256), chunked prefill size (256-8192 tokens), KV cache eviction policy (LRU, LFU, cost-aware), routing policy (round-robin, sticky, cache-aware), and disaggregation strategy (co-located vs. separate prefill/decode)~\cite{distserve,sarathi2023,mooncake} --- yielding a large configuration space. Prior research has shown that tuning configuration choices in accordance with model and workload characteristics can improve throughput by 3-5$\times$ \cite{vidur,distserve}. For instance, Mitra et al. \cite{mitra2025beyond} show that serving Llama-70B with PD disaggregation \cite{splitwise,distserve} can provide 1.8$\times$ throughput improvement over co-location  \cite{taming} for RAG-like, prefill-heavy workloads (16K input tokens, 2K output tokens). Yet for decode-heavy workloads (2K input, 8K output), they find that disaggregation provides minimal benefit ($<$10\% throughput gain) and can even reduce performance due to KV cache transfer overhead and dynamic shift in traffic patterns.

To effectively deploy these systems, operators evaluate these configurations on representative workloads and pick those that satisfy their latency requirements while providing maximal throughput. Evaluating each 
configuration requires around 2-4 hours of profiling to collect statistically significant tail latency characteristics across request load levels ~\cite{vidur}. For a 64-GPU cluster at current cloud pricing (\$2.50-\$7 per GPU-hour for an H100), evaluating just 100 configurations takes 12,800-25,600 GPU-hours, costing \$32,000-\$179,200 USD. These factors make broader exploration of the design space economically and practically infeasible, forcing practitioners to resort to \textit{rule-of-thumb} heuristics.

\subsection{Discrete-Event Simulators}
\label{sec:background:sims}
To reduce the cost of evaluating a configuration, researchers have developed Discrete-Event Simulators (DES) ~\cite{vidur,llmservingsim,apex,frontier,llmservingsim2,tokensim}. These simulators provide a cheap and fast way to model the performance of a serving system without the need for expensive hardware.  These tools operate by manually re-implementing the serving system's control logic -- schedulers, memory allocators, and request routers -- within a simplified simulation framework. A performance model predicts GPU execution time for the batch. The simulator advances its clock by this predicted duration and processes the next event. By eliminating physical GPU execution, simulators achieve up to an order of magnitude speedups over real-time without requiring hardware. 

\subsection{Limitations of Simulators}
\label{sec:des_limitations}

\vheading{Semantic Gap}
The re-implementation effort inevitably introduces subtle differences between the simulator and actual system behavior. Early systems were tractable to model. For instance, Vidur modeled the original vLLM scheduler in $\sim$150 lines of code. Today, production control planes exceed tens of thousands of lines of code \cite{vLLM:github,sglang:github} with complex logic for implementing features like hierarchical prefix caching and disaggregated execution. Capturing the nuanced performance characteristics of these features requires an unwieldy amount of engineering effort. Any deviation in the simulator's logic from the ground truth leads to critical inaccuracies in performance prediction.

\vheading{Framework Specificity}
Simulators also struggle with generality because serving frameworks often make fundamentally different implementation choices for the same feature. For instance, while both vLLM and SGLang implement the batching policy described in Sarathi-Serve \cite{taming} at high-level, they critically differ in their implementation. vLLM adopts both chunked-prefills as well as mixed batching (combining prefill and decode requests in a single batch), whereas SGLang by default doesn't perform mixed batching and resorts to a prefill prioritizing policy. This difference leads to divergent latency characteristics that a generic implementation would fail to capture. We observe such differences in other features too, such as prefix caching, disaggregated execution, etc. For example, Hierarchical cache in vLLM (using LMCache \cite{lmcache:github}) uses a write-through policy, where every cache element written to GPU is also immediately written to lower tiers (CPU, NVMe, etc.). SGLang, on the other hand, adopts a write-through-selective policy that asynchronously copies data only on the first cache hit.

To model this landscape accurately, researchers are forced to build and maintain separate, specialized simulators for each major framework: Vidur \cite{vidur} is modeled after Sarathi-Serve \cite{taming}, and LLMServingSim2.0 \cite{llmservingsim2} is modeled after vLLM. This approach both limits flexibility and increases engineering effort required to construct these simulators.

\vheading{Maintenance Burden}
The engineering overhead of building a DES is exacerbated by the breakneck pace of LLM server evolution. vLLM alone has received over 2,800 commits from hundreds of contributors in the first half of 2025 \cite{vLLM:github}. Each commit has the potential to invalidate the modeling assumptions of a simulator. As a result, simulators like Vidur and LLMServingSim have historically fallen behind shortly after release. There have been attempts to bridge this gap --- Frontier \cite{frontier} and LLMServingSim2.0 \cite{llmservingsim2} have extended the past generation of simulators with modern features like prefix caching and PD disaggregation. However, even these efforts are destined to face the inevitable maintainability challenge as the field progresses.\\\\

\textit{\textbf{Takeaway:} Existing discrete-event simulation approaches face a perpetual maintenance burden. We need a fundamentally new runtime modeling approach that does not  require manual re-implementation of the serving engine logic.}

\section{Time-Accelerated Emulation}
\label{sec:time_accel_emulation}

To address the limitations of discrete-event simulation, we propose a new modeling paradigm: \textit{Time-Accelerated Emulation}. Instead of re-implementing the serving system's logic, this approach allows us to  the \textit{directly} execute unmodified LLM engine, providing high fidelity performance modeling while retaining the speed and cost-effectiveness of simulation. This section outlines the key building blocks that enable this approach.

\subsection{Transparent Device Virtualization}
\label{sec:emulation:virtualization}
A recent work in deep learning training performance modeling, Maya~\cite{maya}, has demonstrated the viability of transparent device emulation. By intercepting CUDA API calls via \texttt{LD\_PRELOAD}, it is possible to create ``virtual devices'' to model the runtime of the workload without need for physical GPUs. Maya achieves this by building a two-step pipeline, first, the workload is executed with the virtual device. This run produces a trace of all the CUDA API invocations made by the workload. A simulator then replays the trace, predicting the runtime of each operation before finally outputting an end-to-end runtime estimate.

Applying this technique to inference serving promises to allow direct execution of unmodified serving system code -- fundamentally eliminating the primary limitations of discrete-event simulation, providing a path toward robust and maintainable runtime modeling for modern LLM serving systems.

\subsection{Temporal Semantics in Online Serving}
Unfortunately, unlike training workloads which are typically offline, inference is an \textit{online} process where request arrival timing fundamentally alters control flow. Ignoring GPU execution time breaks these temporal semantics, leading to incorrect batching decisions.

\todo{Figure would be nice to highlight the conundrum}
For instance, suppose we have two requests, A and B, that both require 500 ms for prefill. Request A arrives at $t=0$, and Request B arrives at $t=200$ ms. In a real system, the scheduler would begin processing A; when B arrives 200 ms later, the GPU is still busy with A. B would be added to the queue, and the scheduler would eventually schedule B once prefill for request A is completed -- perhaps batching B's prefill with A's subsequent decode steps. However, if we simply trace the GPU without accounting for GPU execution in real-time (as in Maya's offline emulation approach), Request A completes instantly at $t \approx 0$. When Request B arrives at $t=200$ms, the system perceives an idle GPU and an empty queue. The temporal overlap is lost, the batching pattern changes completely, and the modeling diverges from reality.

\vheading{A Strawman Approach}
To preserve temporal semantics, the emulator must force the control plane to experience the passage of time. A naive but correct approach is to mimic GPU latency by \textit{sleeping} for the predicted duration of the kernel. If a batch is predicted to take 500 ms, the emulator blocks the CPU thread for 500 ms. This restores fidelity: during the 500 ms sleep for Request A, the wall clock advances. Request B arrives at $t=200$ ms as scheduled, but now finds the worker ``busy'' (sleeping). The scheduler correctly queues B, preserving the system's state dynamics and resource contention exactly as they would appear on real hardware. 

While sleep-based emulation guarantees correctness, it is prohibitively slow. Evaluating a 2-hour trace would require 2 hours of real time. For configuration searches requiring thousands of evaluations, this real-time constraint renders the approach impractical. We refer to this strawman approach as \textit{Sleep-Based Emulation}.

\begin{figure}[t]
    \centering
    \includegraphics[width=0.95\linewidth]{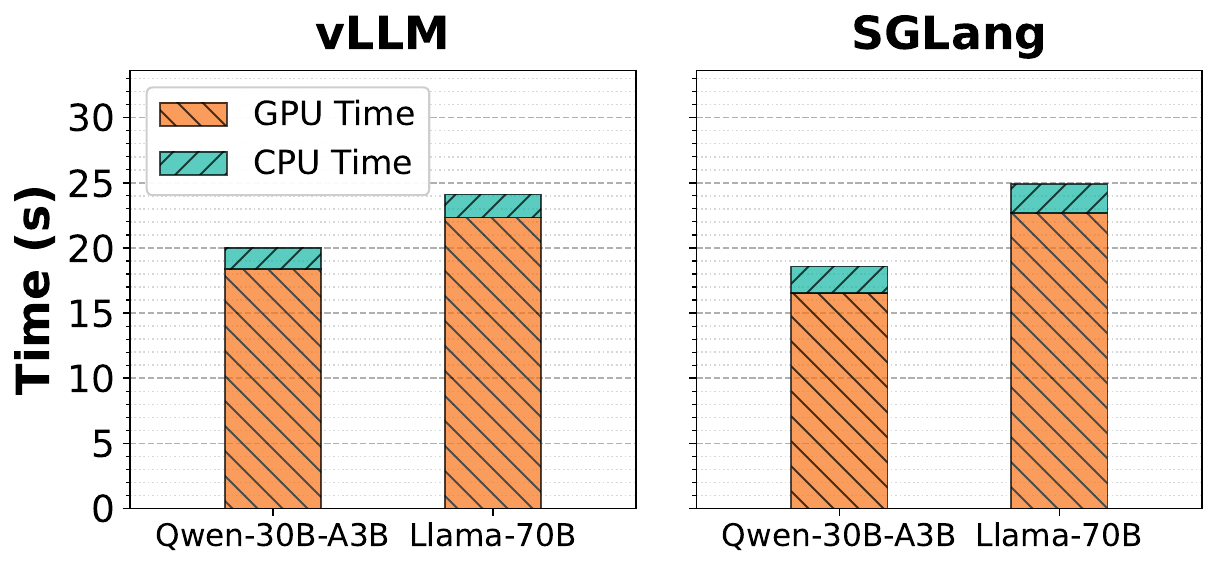}
    \mycaption{CPU vs GPU execution time.}{GPU execution time dominates serving latency. The low CPU overhead (5-10\%) creates the opportunity for order-of-magnitude speedup via time acceleration.}
    \label{fig:cpu-gpu-breakdown}
\end{figure}

\subsection{Key Insight: Time Acceleration}
\label{sec:emulation:enabling-insight}
The strawman emulation approach pays wall-clock cost identical to full GPU execution. However, a closer look at the design of inference engines reveals an opportunity: these systems are engineered to maximize GPU utilization while keeping CPU overhead minimal. \autoref{fig:cpu-gpu-breakdown} quantifies this observation. We process 100 requests in offline mode (all requests available at start) and measure execution time under two conditions -- normal GPU execution and GPU-skip mode where all GPU execution is bypassed via emulation. The CPU time represents the critical-path control-plane overhead, while GPU time captures actual kernel execution. Across both vLLM and SGLang, GPU computation accounts for 90-95\% of total execution time, with CPU overhead remaining minimal regardless of model size or serving framework. Consequently, the system spends the vast majority of wall-clock time simply waiting for GPU computations to complete.

Furthermore, crucially, the control plane's logic does not depend on the \textit{values} computed by the GPU. Schedulers operate on logical abstractions (requests, batches, memory blocks), while GPU workers perform physical computation without making control flow decisions\footnote{Note: In real deployments, the request termination is determined by the generation of EOS tokens. However, for performance modeling purposes, it is a common practice to run each request for a predetermined number of tokens by running specific sampling parameters.}. These properties naturally yield a path to a critical optimization: instead of sleeping during the 85-95\% of time the system is idle, we can potentially \textit{jump} virtual time to accelerate emulation.

\vheading{Tackling Causality}
While the concept of time-jumping is straightforward in principle, implementing it in a distributed serving system introduces a fundamental challenge with respect to causality. In a serving system, multiple independent processes (the batch scheduler, workers, and the request load generator) operate concurrently. Each process has its own view of the next event. For example, a worker may be ready to \textit{jump} 50 ms to simulate batch execution, while the load generator is scheduled to inject a new request in 10 ms.

If the worker unilaterally advances its local clock by 50 ms, it would effectively ``step over'' the arrival of the new request. In the virtual timeline, the request arrives \textit{after} the batch completes, whereas in reality, it would have arrived \textit{during} execution, potentially altering the scheduler's decision for the subsequent batch. This violation of causality leads to incorrect queuing behavior and invalid latency measurements.

\vheading{Time Acceleration in the Literature} This is a classic problem, widely studied in the distributed event simulator literature. Optimistic approaches \cite{virtualtime,timewarp} allow different processes to advance their clocks independently. When a causality violation is detected, the process state is rolled back to the last consistent checkpoint. However, this approach is infeasible for emulation because serving systems cannot be rolled back. In contrast, conservative approaches \cite{distributedsim, asyncdistributedsim, packetcommunication} are more amenable to emulation. These approaches require every process in the system to agree on a common lookahead time within which execution can safely be advanced without violating causality.

While there have been several implementations of conservative time acceleration discussed in literature for discrete-event simulation~\cite{packetcommunication, distributedsim, asyncdistributedsim}, none are directly applicable to emulation. In an event driven simulator, the event loop has direct control over causality of events. In contrast, in emulation systems, all the distributed components operate in real-time, and we need to \emph{transparently} accelerate time without introducing any additional overhead. For instance, the Chandy-Misra-Bryant \cite{asyncdistributedsim,packetcommunication} algorithm requires each simulation worker to send a null message to each peer with the next lookahead time. The simulator needs to be able to control the event scheduling to implement this approach. In this paper, we propose an approach to enable time acceleration without violating causality in emulation systems.

\begin{figure}[t]
\centering
\includegraphics[width=\columnwidth]{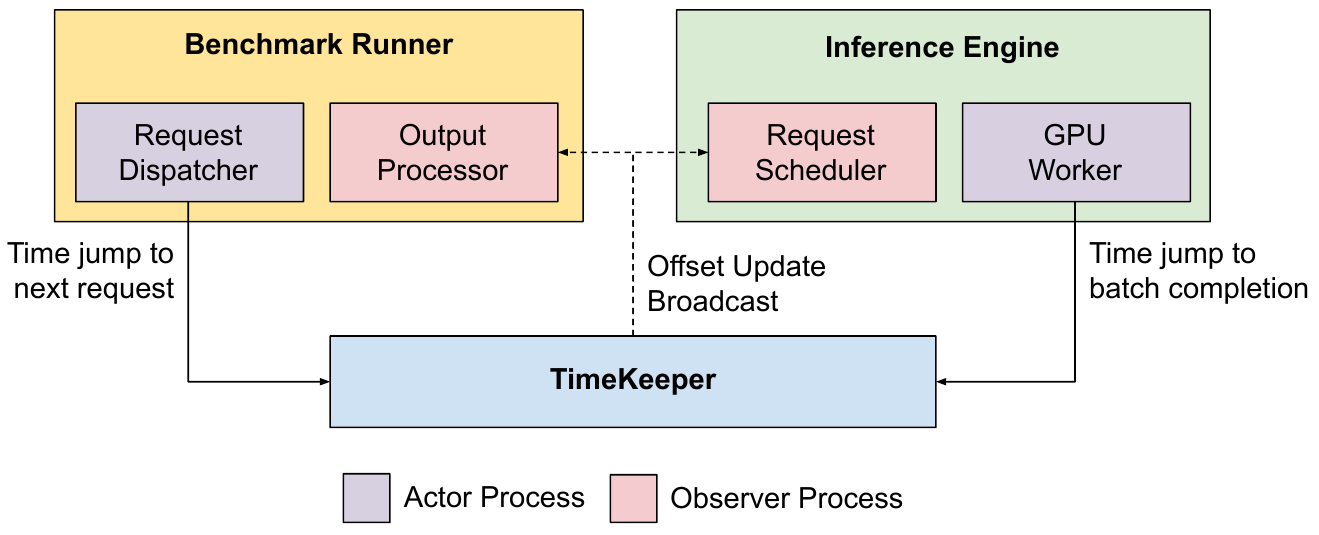}
\mycaption{Coordinating virtual time across processes.}{The benchmark runner and inference engine run as separate processes, each containing actors (blue) that request time jumps and observers (pink) that receive time updates. \timekeeper coordinates jump requests and broadcasts consistent offsets.}
\label{fig:process-model}
\end{figure}

\section{\sysname: System Design}
\label{sec:design}

\sysname\ bridges the gap between parallel distributed event-driven simulation and emulation approaches through \textit{Time-Warp Emulation}. Realizing this paradigm requires resolving the tension between the asynchronous nature of distributed serving systems and the strict causality required for accurate modeling. The rest of this section details the design of \sysname\ and how it addresses challenges related to time virtualization.

\subsection{Overview}
\label{sec:design:overview}

\vheading{Process Model} Running a performance evaluation for a LLM serving system involves two main pieces: a benchmark runner, which submits requests, and the LLM inference engine, which processes these requests~(\Cref{fig:process-model}). These components map to two temporal roles in \sysname{}'s execution model. \textit{\actors} have deterministic plans and know when their future actions will complete. For instance, the benchmark runner's request dispatcher knows when to dispatch each new request (e.g., at t=100ms, t=250ms), and the inference engine's GPU workers know when batch execution will finish. These processes actively drive virtual time forward. \textit{\observers} are purely reactive and respond to events as they occur. The benchmark runner's output processor and the inference engine's request scheduler are two such examples.

\vheading{System Architecture} \sysname{} comprises two main components: the \textbf{Timekeeper}, which synchronizes virtual time across distributed components and manages time jumps to accelerate execution, and the \textbf{\mbox{Device Emulation Layer}}, which transparently intercepts and models GPU operations without requiring physical hardware.

Integrating these pieces with \sysname{} involves patching every \actor in the system to make time jumps when possible. The benchmark runner must be modified to make time jumps between request dispatches. Similarly, the GPU workers of the inference engine need to jump over GPU batch executions. Combined, these take up about 50 lines of code for vLLM and SGLang. Device Emulation Layer is loaded via \texttt{LD\_PRELOAD} to intercept any device management calls separate from GPU kernel dispatches. When execution begins, \actors submit concurrent time jump requests to the \timekeeper --- the benchmark runner requesting a jump to the next request arrival, GPU workers requesting jumps over their batch execution durations. The Timekeeper computes the minimum target across all \actors, advances the global virtual clock to that point, and broadcasts the update. \actors whose targets have been reached return immediately. \actors with further targets remain blocked and resubmit in subsequent barrier rounds. Observers query virtual time asynchronously to timestamp events without participating in this coordination. %

\subsection{\timekeeper}
\label{sec:design:timekeeper}
The \timekeeper is a service that manages virtual time across connected clients. It exposes a \textsc{TimeJump($\Delta t$)} API that allows clients to request time jumps and receive clock updates. We classify clients of the \timekeeper as either \actors or \observers. \actors are active drivers of the simulation that process operations with predictable durations and must actively request virtual time advancement. \observers are reactive components that consume virtual time to timestamp events but do not block its progression. This distinction allows \sysname\ to minimize coordination overhead by requiring synchronization only from \actors, while enabling integration with serving frameworks through minimal code changes.

\begin{figure}[t]
\centering
\includegraphics[width=\columnwidth]{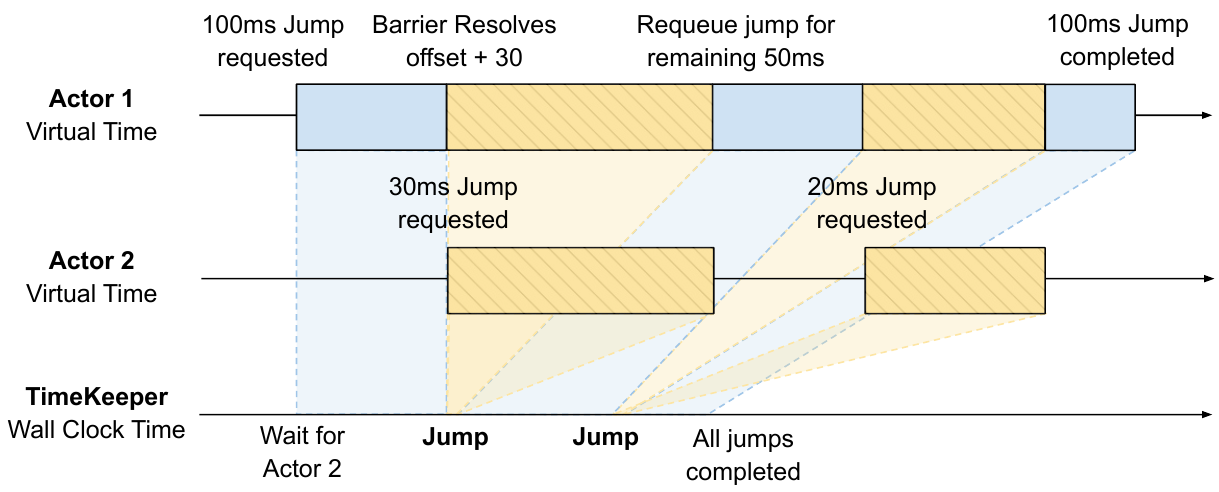}
\mycaption{Barrier-based time synchronization.}{When actors request different jump durations, \timekeeper{} advances to the minimum safe offset. Here, Actor 1 requests a 100ms time jump at 
$t=0$. Before performing the jump, \timekeeper{} waits for all actors to publish their safe offsets. When Actor 2 requests a 30ms jump at $t=20$, the barrier resolves and \timekeeper{} advances time by 30ms. Actor 1 still has 50ms remaining, so it requeues a jump request for the remainder. Actor 1's original 100ms jump is thus processed in two chunks, with total wall clock time of just 50ms.}
\label{fig:timejump-resubmit}
\end{figure}

To ensure scalability, the architecture separates the \textit{request path} from the \textit{update path}. \actors\ submit time jump requests individually to the \timekeeper via a reliable channel (fan-in). The \timekeeper disseminates clock updates to all clients simultaneously via a broadcast channel (fan-out). This asymmetry allows the \timekeeper to update hundreds of distributed workers with constant serialization cost per round, preventing bottlenecks during high-frequency barrier resolutions.

\subsubsection{Virtual Time Protocol}
\label{sec:design:timekeeper:time-protocol}

Coordinating virtual time across distributed processes requires a protocol that preserves causality without controlling process execution. This section presents \sysname's barrier-based protocol and analyzes its correctness properties.

\vheading{Design Constraints}
Adapting virtual time to real-time emulation introduces three constraints absent in traditional discrete-event simulators:

\begin{itemize}[leftmargin=*,nosep]
\item \textit{\textbf{No rollback.}} Optimistic approaches like Time Warp~\cite{virtualtime, timewarp} speculatively advance time and rollback on causality violations. Since we execute real system code with irreversible side effects---network messages sent, data structures modified---this approach is infeasible.

\item \textit{\textbf{No event scheduling control.}} Conservative protocols like Chandy-Misra-Bryant~\cite{asyncdistributedsim} assume the simulator controls event processing order. Our processes run asynchronously in wall-clock time; the protocol must coordinate time advancement without controlling execution.

\item \textit{\textbf{Graceful degradation.}} When coordination fails (stragglers, network delays), the system must remain correct, potentially sacrificing speed but never producing invalid results.
\end{itemize}

\begin{algorithm}[t]
\caption{Client: TimeJump($\Delta t$)}
\label{alg:client-time-jump}
\begin{algorithmic}[1]
\Require $\Delta t > 0$: virtual time to advance (ms)
\State $t_{target} \gets \textsc{GetVirtualTime}() + \Delta t$
\Statex $\triangleright$ \textit{Compute absolute target once}
\While{$\textsc{GetVirtualTime}() < t_{target}$}
    \State \textsc{SendTimeJumpRequest}$(t_{target})$
    \Statex \quad $\triangleright$ \textit{Request advance to target}
    \State \textsc{WaitForAck}()
    \State $t_{remaining} \gets t_{target} - \textsc{GetVirtualTime}()$
    \If{$t_{remaining} > 0$}
        \State \textsc{WaitForClockUpdate}$(t_{remaining})$
        \Statex \qquad $\triangleright$ \textit{Block until broadcast or timeout}
    \EndIf
\EndWhile
\end{algorithmic}
\end{algorithm}

\vheading{Virtual Time Representation}
\sysname\ maintains a global virtual clock as an offset from wall-clock time:
\begin{equation}
t_{virtual} = t_{wall} + \mathit{offset}
\end{equation}
Initially $\mathit{offset} = 0$, so virtual time equals wall time. As \actors\ request time jumps, the \timekeeper increases the offset, causing virtual time to advance faster than wall time. This representation allows \observers\ to query virtual time without coordination---they simply read the current offset and add wall time.

\vheading{Barrier-Based Coordination}
When an \actor\ calls \textsc{TimeJump}($\Delta t$), it computes its target virtual time and sends a request to the \timekeeper. The \timekeeper collects requests from all \actors, then advances virtual time to the \textit{minimum} requested target. This minimum-advancement rule ensures no \actor\ jumps past its intended time, preserving causality. For example, if one worker needs 50ms to process its batch while another needs 10ms, advancing each independently would break causality: an \actor\ at virtual time $t=1000$ could observe events from an \actor\ still at $t=500$. Algorithm \autoref{alg:server-barrier} shows the server-side protocol.

A single \textsc{TimeJump} call may span multiple barrier rounds. Consider two workers: $W_A$ calls \textsc{TimeJump}(50), $W_B$ calls \textsc{TimeJump}(10). The \timekeeper computes $t_{min}$ from $W_B$'s target and advances by 10ms. $W_B$'s call returns. $W_A$ remains in its loop---virtual time has advanced but not to $W_A$'s target---and re-requests in the next barrier round. \Cref{alg:client-time-jump} shows the client-side protocol. The client computes its target time once (line 1), then loops until virtual time reaches this target. Each iteration sends the target to the \timekeeper, receives acknowledgment, and waits for a clock update broadcast with a timeout equal to the remaining virtual time needed.

\begin{algorithm}[t]
\caption{Server: ProcessTimeJumpRequests()}
\label{alg:server-barrier}
\begin{algorithmic}[1]
\Require $numActors$: number of registered actors
\State $pending \gets \emptyset$, $offset \gets 0$
\Statex $\triangleright$ \textit{Initialize pending requests and virtual offset}
\While{\textbf{true}}
    \State $(c, t_{target}) \gets \textsc{ReceiveRequest}()$
    \State $pending[c] \gets t_{target}$
    \Statex \quad $\triangleright$ \textit{Store request for client $c$}
    \If{$|pending| = numActors$}
        \Statex \qquad $\triangleright$ \textit{All actors at barrier---advance virtual time}
        \State $t_{min} \gets \min\{t : t \in pending\}$
        \Statex \qquad $\triangleright$ \textit{Minimum target preserves causality}
        \If{$\textsc{GetWallTime}() < t_{min}$}
            \State $t_{wall} \gets \textsc{GetWallTime}()$
            \State $offset \gets \max(offset, t_{min} - t_{wall})$
            \State \textsc{BroadcastClockUpdate}$(offset)$
            \Statex \qquad\quad $\triangleright$ \textit{Notify all clients}
        \EndIf
        \State $pending \gets \emptyset$
        \Statex \qquad $\triangleright$ \textit{Reset for next round}
    \EndIf
\EndWhile
\end{algorithmic}
\end{algorithm}

\vheading{Graceful Degradation}
The timeout in \Cref{alg:client-time-jump} ensures correctness under failures. If the broadcast never arrives---due to network issues or a stalled \actor---the wait times out after $t_{remaining}$ wall-clock milliseconds. At that point, wall time has advanced by $t_{remaining}$, and since $t_{virtual} = t_{wall} + \mathit{offset}$, virtual time has also advanced. The loop condition succeeds, and \textsc{TimeJump} returns.

The operation completes correctly but at wall-clock speed. In the worst case, \sysname\ degenerates to sleep-based emulation---slow but never incorrect. This property guarantees \sysname\ cannot produce invalid results or deadlock; it can only lose acceleration.

\vheading{Handling Message Jitter}
Real-time emulation must account for network delays and CPU scheduling jitter. Consider this scenario: the inference server generates a token at wall time $t_w{=}100ms$ and virtual time $t_v{=}3400ms$. The benchmark runner receives it promptly and records $t_v{=}3400ms$. The server then processes the next batch (20ms predicted), updating virtual time to $t_v{=}3420ms$ at $t_w{=}101ms$. Due to network delay, the runner reads the next token at $t_w{=}102.5ms$. Meanwhile, the server completes another batch and updates to $t_v{=}3450ms$ at $t_w{=}102ms$. When the runner queries virtual time at $t_w{=}102.5ms$, it observes 3450ms instead of 3420ms---an inaccurate latency measurement.

Attaching virtual timestamps to messages as soon as they are produced would eliminate this ambiguity. While accurate, this approach also requires significant changes (a few hundred lines) to the serving engine. For easier integration, \sysname also offers a \textit{bounded-jitter model}: assuming messages are received within $J$ time of transmission, inserting a $J$-duration cooldown between consecutive time jumps prevents \observers\ from reading stale virtual times. We find that $J \approx 500\mu s$ suffices in most practical settings.

\begin{figure*}[t]
\centering
\includegraphics[width=\textwidth]{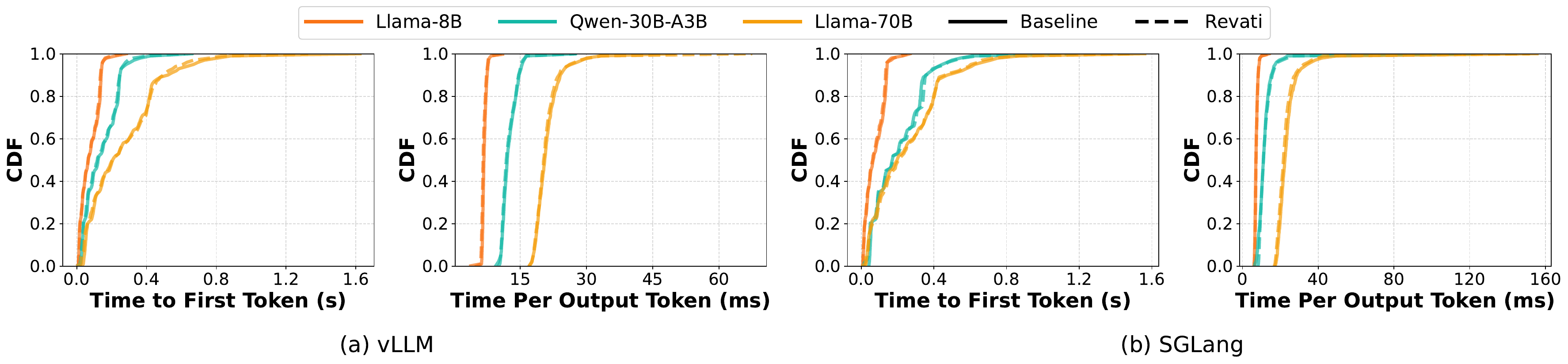}
\mycaption{End-to-end accuracy.}{\sysname{} (dashed) closely matches real execution (solid) for both TTFT and TPOT distributions across three model configurations. Prediction error remains below 5\% even at tail.}
\label{fig:eval:e2e:accuracy}
\end{figure*}

\begin{figure}[t]
\centering
\includegraphics[width=0.6\linewidth]{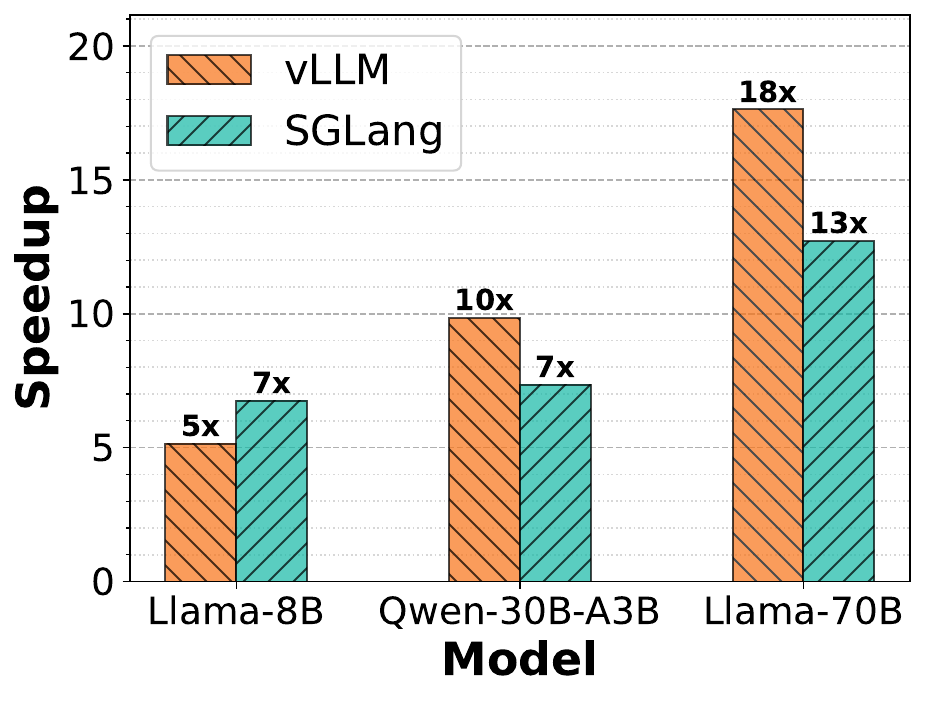}
\mycaption{End-to-end speedup.}{\sysname{} achieves 5-17$\times$ speedup on vLLM and 6-12$\times$ on SGLang compared to real on GPU execution.}
\label{fig:eval:e2e:speedup}
\end{figure}

\subsection{Device Emulation Layer}
\label{sec:design:emulation-layer}

Serving frameworks are deeply integrated with CUDA and NCCL. Modifying all call sites would require invasive per-framework changes, undermining \sysname's transparency goal. The Device Emulation Layer adapts Maya's~\cite{maya} transparent device virtualization to intercept and emulate CUDA API calls for inference workloads. Unmodified framework code executes while believing it has access to target hardware. In addition to maintaining transparency, this approach enables evaluation at scales beyond available resources. For example, a researcher who wants to perform evaluations on a 128-GPU H200 cluster, but doesn't have access to such hardware can simply configure \sysname{} to emulate the desired hardware.

However, performing emulation in real-time (as opposed to Maya's offline approach) introduces some unique challenges that we discuss below.

\vheading{Preserving Distributed Dependencies}
In distributed serving (pipeline parallelism, tensor parallelism), NCCL collectives enforce causal dependencies---stage $i{+}1$ cannot proceed until stage $i$ completes \texttt{ncclSend}. Without actual GPU execution, \sysname\ must block emulated stage $i{+}1$ until emulated stage $i$ reaches the corresponding send. We convert NCCL collectives into barrier synchronization points across participating workers, preserving temporal ordering without data transfer.

\vheading{Split-State Memory Model}
Unlike training workloads, serving systems often use GPU channels for control-plane communication. SGLang, for instance, broadcasts batch composition via NCCL. Pure black-box emulation would break this because the CPU expects valid data in these buffers. To tackle this, \sysname\ bifurcates allocations into two categories. Metadata buffers -- small allocations (below 4MB by default) potentially used for control decisions -- are backed by real host memory and their operations execute faithfully. Compute buffers -- large allocations for KV caches and weights -- receive virtual pointers with no physical backing and their operations become no-ops with durations estimated by the runtime predictor. This model relies on compute buffers never being read by the CPU. To enforce this invariant, \sysname raises a fatal exception if the application attempts to read from a virtual compute buffer rather than returning garbage values. A successful emulation run thus guarantees the control plane never operated on phantom data.

\subsection{Runtime Prediction}
\label{sec:runtime-prediction}

\sysname\ provides a pluggable interface supporting both analytical and profiling-based runtime predictors. By default, we extend Vidur's~\cite{vidur} operator-level models with support for Mixture-of-Experts routing, fused attention variants, and optimized all-reduce collectives. The predictor accepts batch composition (prefill count, decode count, sequence lengths) and target hardware spec, returning an estimated duration.

\section{Implementation}
\label{sec:design:implementation}

The \timekeeper server and client library comprise ${\sim}800$ lines of C++, with Python bindings for framework integration. The device emulator adds ${\sim}6000$ lines of C++. LLM server patches total fewer than 50 lines each for vLLM and SGLang. The messaging layer uses ZeroMQ for low-latency communication. The \timekeeper server employs a multi-threaded architecture: a dedicated I/O thread handles serialization and socket operations, and a background thread manages the barrier state and time-jump logic. This separation ensures high-frequency requests do not block barrier resolution.

\section{Evaluation}
\label{sec:evaluation}

In this section, we present a comprehensive evaluation of \sysname to demonstrate its fidelity in simulating LLM inference performance.

\todo{add section refs}
\begin{itemize}[itemsep=0pt]
    \item \textbf{Accuracy}: How accurately does \sysname predict end-to-end inference latency across varying model sizes, and deployment configurations?
    \item \textbf{Efficiency}: Can \sysname effectively accelerate evaluations compared to the strawman  approach?
\end{itemize}

\subsection{Experimental Setup}

\vheading{Models and Inference Systems}
We conduct profiling and simulation across a spectrum of model sizes to ensure generalization. Specifically, we evaluate the performance of our system across 2  dense models (\llamaS \& \llamaL),  and one sparse model (\qwenMM). We set the tensor parallel degrees to one and four for \llamaS and \llamaL  respectively. We run \qwenMM with expert parallel degree of two. To evaluate the generality of the system, we perform experimentation across two different serving engines -- vLLM \cite{vLLM:github} and SGLang \cite{sglang:github}. To ensure a consistent comparison, we enable chunked prefill with mixed batching across all settings with chunk size 512.

\vheading{Hardware}
We perform our evaluations on a machine with 4 H200 GPUs featuring a fully connected NVLink fabric with a 128 core AMD 9334 EPYC CPU and 756GB memory.

\begin{figure}[t]
    \centering
    \includegraphics[width=\linewidth]{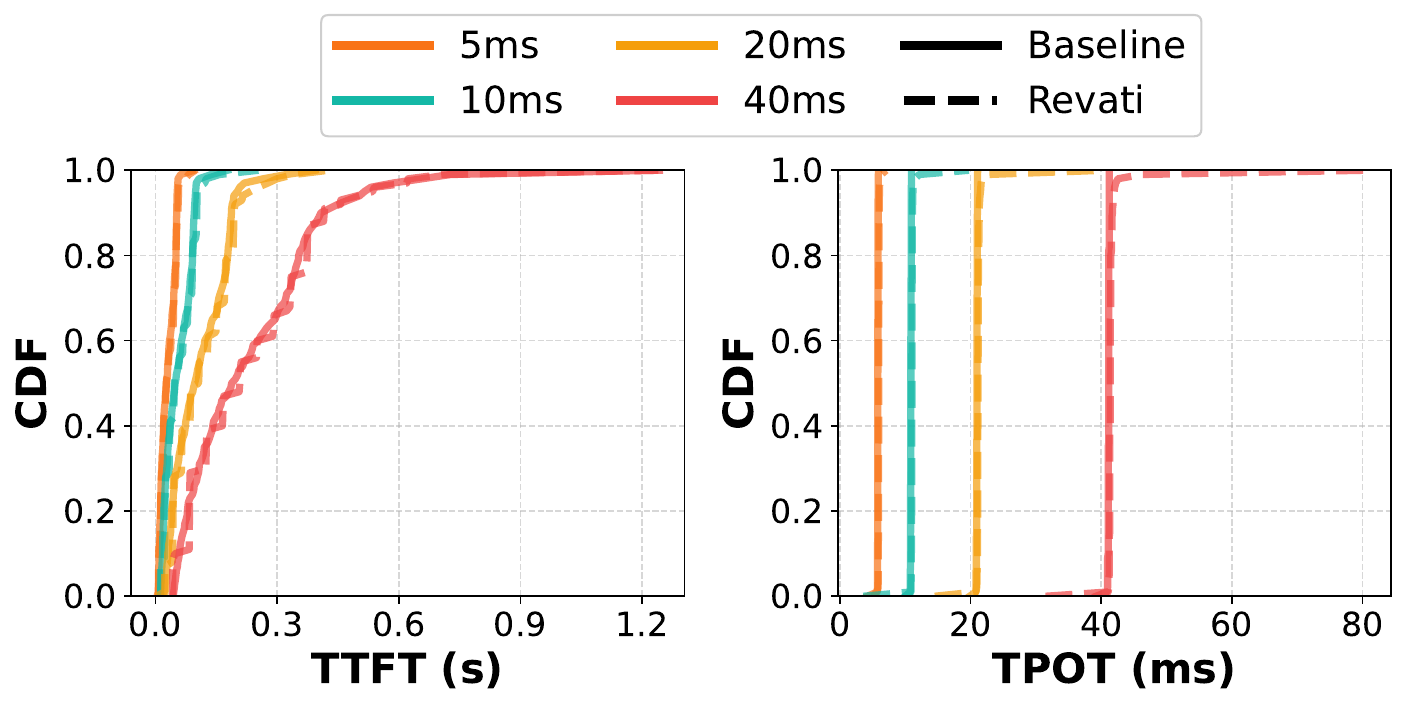}
    \mycaption{Accuracy with Varying batch durations.}{\sysname{} (dashed) matches sleep-based emulation (solid) across all batch durations.}
    \label{fig:eval:ablation:batch}
\end{figure}

\begin{figure}[t]
    \centering
    \includegraphics[width=0.9\linewidth]{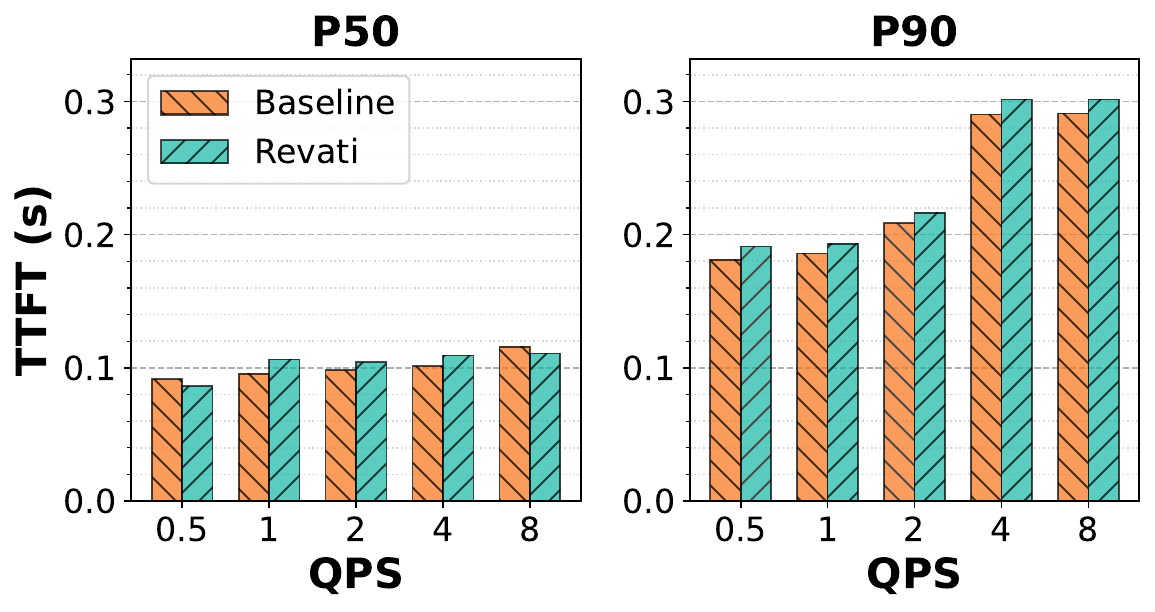}
    \mycaption{Accuracy with varying request rate.}{\sysname{} accurately models both the median and tail latency trends across varying request arrival rates.}
    \label{fig:eval:ablation:qps}
\end{figure}

\begin{figure}[t]
    \centering
    \includegraphics[width=0.8\linewidth]{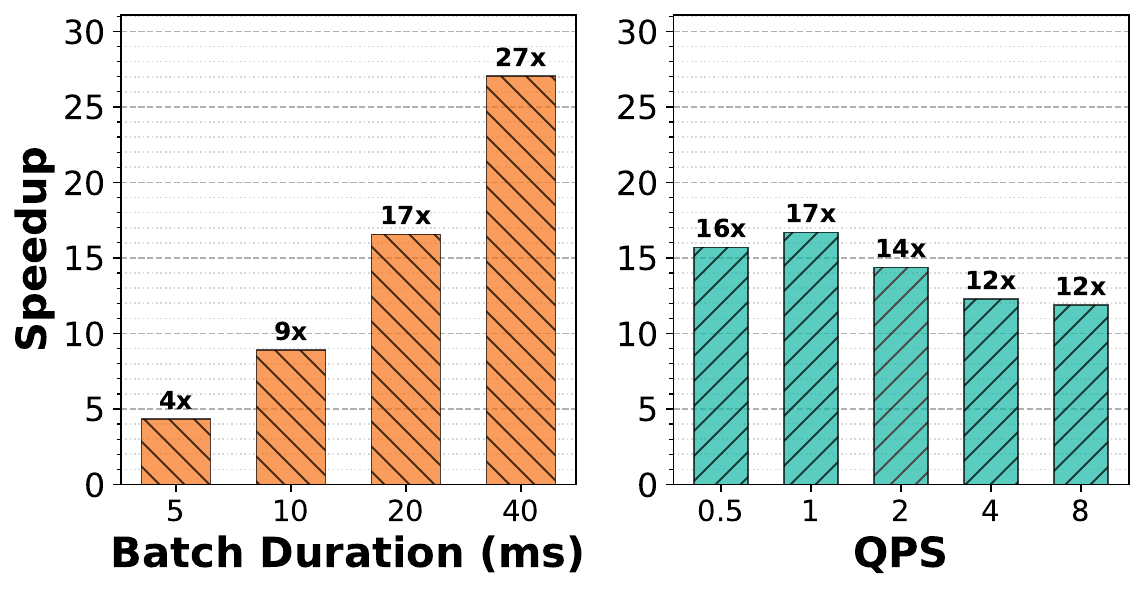}
    \mycaption{Speedup with varying parameters.}{\sysname consistently maintains high speed up across varying batch durations and request rates.}
    \label{fig:eval:ablation:speedup}
\end{figure}

\subsection{End-to-end Evaluations}
We evaluate \sysname{}'s accuracy and speedup on end-to-end inference benchmarks across two production serving engines (vLLM and SGLang), three model configurations (\llamaS{}, \llamaL{}, and \qwenMM{}), using the ShareGPT dataset and Poisson arrival.

\vheading{Accuracy} 
\autoref{fig:eval:e2e:accuracy} compares the latency distributions predicted by \sysname{} (dashed) against real execution (solid). Across all model sizes and both serving engines, \sysname{} closely tracks the baseline Time-to-first-token (TTFT) and Time-per-output-token (TPOT) distributions. The predicted latencies match near identically across the CDF with median prediction error below 5\%. By directly executing serving engines, \sysname is able to accurately model minute details in the scheduling policies of these systems. We observe that though both the systems exhibit similar TTFT performance, SGLang suffers from a significantly worse tail when it comes to decode TPOT -- this is because, SGLang does not perform mixed batching by default, though it performs chunked prefills.

\vheading{Speedup} As shown in \autoref{fig:eval:e2e:speedup} \sysname{} achieves over an order of magnitude speedups across both vLLM and SGLang over real execution due to its time acceleration protocol. Note that as models get larger and the GPU execution time increases, we obtain higher speed ups with peak performance for \llamaL.

\subsection{Ablation Experiments}

To understand how workload characteristics affect \sysname{}'s performance, we conduct various ablation studies.

\vheading{Varying Batch Durations} First, we evaluate how accurately and efficiently \sysname can operate if the model size and hardware changes. We mimic this by using static batch time predictions of varying durations between 5 and 40 ms. We compare our performance against the naive sleep-based approach, where GPU workers simply sleep for the batch duration instead of performing time jumps.

As shown in \autoref{fig:eval:ablation:batch}, \sysname{} produces highly accurate TTFT and TPOT distributions with less than 5\% error.  As we increase the batch duration, we obtain higher speedup --- going up to 27$\times$ at batch runtime of 40ms. This is a direct consequence of \sysname{}'s ability to skip over GPU compute wait times.

\vheading{Varying Arrival Rates} Another interesting aspect of the workload which could directly affect the accuracy and efficacy of virtual time advancement protocol is the request arrival rate. \autoref{fig:eval:ablation:qps} shows \sysname's performance on request arrival rate varying between 0.5 and 8 queries per second (QPS). To exclude any errors introduced due to the runtime predictor, we run this experiment with a fixed batch time of 20 ms and compare against the sleep-based baseline. The system maintains less than 5\% error in TTFT predictions across the board.

As the arrival rate increases, we observe slight reduction in speed up. This is because at higher arrival rates the system operates with larger batch sizes, which results in higher CPU work for batch formation and bookkeeping. However, since this overhead is relatively small (typically < 200$\mu$s) compared to batch execution time, \sysname sustains over an order of magnitude speedup even at high request load.

\section{Related Work}
\label{sec:related-work}

\vheading{Virtual Time in Distributed Simulation}
Jefferson's Time Warp~\cite{virtualtime, timewarp} pioneered optimistic virtual time: processes advance speculatively and rollback on causality violations. Chandy-Misra ~\cite{asyncdistributedsim, packetcommunication} introduced conservative protocols where processes exchange lookahead information to advance safely. Both assume the simulator controls event scheduling. \sysname\ operates under a different constraint: processes run asynchronously in wall-clock time with real side effects that cannot be rolled back. Our barrier-based protocol adapts conservative principles to this setting, using minimum-target advancement to preserve causality while timeouts ensure graceful degradation to real-time execution.

\vheading{Time Dilation for Network Emulation}
Time dilation~\cite{gupta2006infinity, modelnet, diecast} slows virtual time to make physical hardware appear faster—a 1~Gbps NIC can emulate 10~Gbps by running applications at $10\times$ slower virtual time. Tools like DieCast~\cite{diecast}, ModelNet~\cite{modelnet}, and Mininet~\cite{mininet} use this technique for network testing at scale. \sysname\ inverts this relationship: rather than slowing time to amplify hardware capability, we accelerate time to skip over GPU computation. Both share the insight that applications can operate in virtual time decoupled from wall-clock time, but the mechanisms differ substantially. Network emulators intercept system calls and scale delays; \sysname\ coordinates distributed processes through explicit barrier synchronization to maintain causality across time jumps.

\section{Discussion}
\label{sec:discussion}

\vheading{Towards Fully Transparent Emulation}
\sysname\ requires a minimal (${\sim}$50 line) patch per framework to invoke time jumps. However, this introduces a natural question: could we eliminate this requirement entirely through kernel-level interception, as Maya does for training? However, there are two challenges with this approach. First, modern inference systems have framework-specific implementations for certain complex operations like flash attention, MoE routing. Predicting their runtime from kernel API signatures alone would require complex reverse-engineering of framework internals -- introducing maintenance burden that \sysname\ aims to avoid. More recently, there have been attempts to directly estimate the runtime of GPU kernels from their source code and input parameters \cite{neusight, ithemal} -- this direction has the promise of allowing fully transparent emulation in future.

\vheading{Speed of Light Execution}
The systems community has been striving hard to minimize the CPU overheads in serving systems to maximize the GPU utilization. This is becoming increasingly important as GPUs become faster and more capable over generations. Recently, there have been multiple efforts to implement highly efficient executions in C++ and Rust \cite{tgi, llamacpp} instead of the traditional Python implementations. Existing frameworks like SGLang have also started to recently migrate critical segments of their codebase like the prefix cache tree to C++ to achieve this goal \cite{sglang:github}. Interestingly, this raises the question: what maximum acceleration can we achieve with \sysname? As the emulation speeds up, we expect to encounter more interesting challenges and bottlenecks in the system -- for instance, the tokenization/detokenization steps could add non-trivial overheads etc. We hope to explore these challenges in future work.

\section{Conclusion}
\label{sec:conclusion}
The rapid evolution of LLM serving systems has rendered traditional discrete-event simulation unsustainable. The gap between a simulator’s model and frontier-grade serving systems is simply too expensive to bridge manually. \sysname\ eliminates this gap entirely by cheap and accurate runtime modeling directly on top of original serving frameworks. We show that the transparent device emulation layer, combined with the barrier-based virtual time protocol, can enable accurate runtime modeling (${<}5\%$ error) while concurrently achieving an order of magnitude speedup over physical execution.

\bibliographystyle{plain}
\bibliography{all}

\end{document}